# Probing mesoscopic nonlocal screening in van der Waals heterostructures with polaritons

Xuezhi Ma[1,2*†], Zhipeng Li[1†], Ruihuan Duan[3†], Zeyu Deng[4†], Hao Hu[5†], Mengting Jiang[1], Yueqian Zhang[6], Xiaoyuan He[6], Qiushi Liu[7], Qiyao Liu[1], Yuan Ma[8], Fengxia Wei[1], Jiayu Shi[3], Chunqi Zheng[4], Guangwei Hu[3], Ping Koy Lam[1,2], Chengwei Qiu[4], Yu Luo[5*], Zheng Liu[3*], Qian Wang[1*]

**Affiliations:**

[1] Institute of Materials Research and Engineering (IMRE), Agency for Science, Technology and Research (A*STAR), 2 Fusionopolis Way, Innovis, #08-03, Singapore 138634, Republic of Singapore

[2] Quantum Innovation Centre (Q. InC), Agency for Science, Technology and Research (A*STAR), 4 Fusionopolis Way, Kinesis, #05, Singapore 138635, Republic of Singapore

[3] School of Materials Science and Engineering, Nanyang Technological University, Singapore 639798, Singapore

[4] Department of Materials Science and Engineering, National University of Singapore, Singapore 117575, Singapore

[5] National Key Laboratory of Microwave Photonics & College of Electronic and Information Engineering, Nanjing University of Aeronautics and Astronautics, Nanjing 211106, China

[6] School of Electrical and Electronic Engineering, Nanyang Technological University, 50 Nanyang Avenue, Singapore 639798, Singapore

[7] Shanghai Institute of Optics and Fine Mechanics, Chinese Academy of Sciences, Shanghai, 201800, China

[8] Department of Mechanical Engineering, The Hong Kong Polytechnic University, Hong Kong 999077, China

[†]These authors contributed equally to this work.

[*]Corresponding author. Email: ma_xuezhi@a-star.edu.sg (X.M.); yu.luo@nuaa.edu.cn (Y.L.); z.liu@ntu.edu.sg (Z.L.); wang_qian@a-star.edu.sg (Q.W.)

**Abstract**

Predictive optical modelling of van der Waals (vdW) heterostructures is critical for meta-optics, near-field photonics and quantum technologies. At their buried interfaces, charge transfer and spatially extended screening challenge local descriptions based on layer-by-layer stacking of fixed permittivity tensors. However, such nonlocal corrections have been established mainly for plasmonic systems at ångström-nanometre scales and are often assumed negligible on optical-wavelength scales. Here we challenge this view by uncovering a mesoscopic nonlocal screening regime, extending up to ~140 nm, at buried charge-transfer interfaces in transition-metal dichalcogenide/α-molybdenum trioxide (TMDC/α-MoO$_3$) phonon-polaritonic heterostructures. Using phonon polaritons as an ultrasensitive probe, we quantify charge transfer from polariton-wavelength shifts and find a thickness-independent saturated response as α-MoO$_3$ is thinned. Rather than merely complicating optical modelling, this nonlocal saturation turns a design-level correction into an opportunity by yielding a transferable cross-material metric. Across more than 120 devices, this metric scales linearly with the work-function difference between the TMDC and α-MoO$_3$. We further identify a lattice-mismatch-set energy threshold for charge transfer, revising Anderson-type band alignment for vdW interfaces.

**Main**

Van der Waals (vdW) heterostructures, assembled by stacking atomically thin layers with minimal constraints on lattice matching[1-3], provide a versatile platform for electronic and optical engineering and have enabled rapid progress across devices and phenomena, from microelectronics and optoelectronics to near-field optics, nonlinear nanophotonics and quantum technologies[4-12]. Realizing these opportunities increasingly relies on predictive optical modelling to translate microscopic material responses into designable behaviour of buried interfaces and guided or surface-confined modes. A common working assumption is that each constituent can be assigned a thickness-independent permittivity tensor, so that heterostructures can be modelled as simple stacks of fixed-parameter layers[3,7].

However, in practical vdW stacks, neighbouring layers can be electronically coupled by interfacial charge transfer and dipole formation, so the optical response of each constituent is no longer an intrinsic, thickness-independent tensor that can be treated as a fixed permittivity[13]. Moreover, the associated screening charge and lattice polarization are not confined to an idealized boundary but extend over a finite depth, rendering a purely local, layer-by-layer permittivity description internally inconsistent[14]. A well-established route beyond such local modelling is nonlocal electrodynamics, which explicitly accounts for spatially extended induced charge[14,15]. In metallic nanostructures, nonlocal corrections are typically invoked to capture finite screening and electron spill-out or tunnelling across sub-nanometre gaps, which smear induced charge over ångström-to-nanometre scales[15-17]. Until now, nonlocality has therefore largely been treated as a nanometre-scale refinement for extreme near-field problems and is routinely neglected on optical-wavelength scales[18-20].

Here we reveal a mesoscopic nonlocal electrostatic regime at buried charge-transfer interfaces in transition-metal dichalcogenide/α-molybdenum trioxide (TMDC/α-MoO$_3$) polaritonic heterostructures. By sweeping the α-MoO$_3$ thickness, we observe a crossover from a depletion-like thickness dependent to a thickness-independent saturation plateau below a critical α-MoO$_3$ thickness of up to ~140 nm. This result shows that nonlocal screening can persist on mesoscopic scales, well beyond the atomic-length regime usually associated with nonlocal corrections and therefore cannot be ignored in local-response optical modelling. We access and quantify this regime by exploiting the high sensitivity of polaritons, especially surface confined phonon polaritons (PhPs), to buried charge-transfer electrostatics[7,13,21-26]. Using scattering-type scanning near-field optical microscopy (s-SNOM)[27-35], we read out charge transfer from the TMDC to α-MoO$_3$ through the shift of the PhPs wavelength. The readout defined as $\lambda_1/\lambda_2 - 1$ (Fig. 1a, b), where $\lambda_1$ and $\lambda_2$ denote the PhPs wavelengths with and without coverage by a TMDC layer, respectively.

Unexpectedly, this nonlocal saturation is not only a nonlocal correction. It also creates a practical calibration regime. Once the polaritonic host α-MoO$_3$ reaches its intrinsic saturation state, the near-field readout becomes thickness-independent and can therefore be compared directly across interfaces. The saturated readout serves a universal metric across more than 120 devices and reveals systematic trends in interfacial charge transfer that are not accessible from isolated case studies. More interestingly, we find that the saturated readout can be linearly modulated by the TMDC-α-MoO$_3$ work function difference ($\Delta\Phi$), establishing a quantitative link between the polaritonic readout and the driving force for interfacial charge transfer. Taken together, these results show that the Anderson-type vacuum-level alignment requires a vdW-specific correction against lateral slide or twist within the range studied[8,36-38].

**Observation of a mesoscopic nonlocal screening saturation**

The polaritonic host α-MoO$_3$ has a low electronic density of states and weak free-carrier screening, making its interfacial electrostatics particularly susceptible to redistribution over an extended depth (as characterised by density functional theory, DFT; see Extended Data Fig. 1 and Supplementary Section 4). We therefore test whether the screening is confined to an atomically thin boundary or instead involves a mesoscopic host volume by systematically varying the α-MoO$_3$ thickness and tracking the buried charge-transfer electrostatics via the polariton readout. To minimize confounding factors, including local defects, spatial inhomogeneity and flake-to-flake variability, we sequentially transfer the same TMDC flake (25-nm-thick WSe$_2$) onto eight α-MoO$_3$ substrates with thicknesses from 66.4 to 293.3 nm using a capillary-force-assisted clean-stamp protocol (Fig. 1c, see Supplementary Figs. 6 for details) [39,40]. Each transfer is aligned to the same α-MoO$_3$ edge region so that the imaging geometry remains comparable while only the effective screening volume is tuned.

The measured readout exhibits two distinct regimes as a function of α-MoO$_3$ thickness (Fig. 1d). For sufficiently thick α-MoO$_3$, the response decreases approximately as $1/t$, consistent with a local-response depletion picture in which the transferred charge is accommodated within the host and the interfacial perturbation is diluted as the screening volume increases (see Supplementary Note 11). In contrast, below a critical thickness the response approaches a plateau and becomes thickness independent, indicating saturation of the interfacial screening response. The transition occurs on a mesoscopic length scale of order 100 nm, far beyond the ångström-nanometre regime where nonlocal effects are usually considered. This thickness-independent saturation is not captured by a fixed-permittivity local stacking model (dashed curve in Fig. 1d), pointing instead to a nonlocal regime in which screening charge redistributes over a finite depth that evolves with the host thickness.

**Feibelman *d*-parameter framework for charge screening saturation**

We capture this thickness-dependent breakdown of local response using the Feibelman *d*-parameter formalism, which builds up an interfacial electronic length scale through mesoscopic surface-response functions[14,41-43] (a full theoretical derivation is provided in Supplementary Section 1). In this framework, the key quantity is the out-of-plane $d_\perp$, defined as the first moment (centroid) of the induced charge density relative to the interface plane (Fig. 1e), and it can be incorporated into macroscopic Maxwell electrodynamics through modified mesoscopic boundary conditions. As α-MoO$_3$ is thinned into the saturation regime, the induced-charge centroid shifts towards the TMDC, corresponding to a rapid increase in the effective $d_\perp$ extracted from the data (Fig. 1f), signalling that the screening response has entered a saturated, thickness-independent regime. In this nonlocal regime, the screening response is displaced away from the α-MoO$_3$ bulk, so the PhPs readout becomes pinned and falls below the prediction of a purely local fixed-permittivity stacking model.

To connect this nonlocal response back to an intuitive local-stack picture, we map it onto an effective participating TMDC thickness, $t_{\text{eff}}(t),$ defined as the thickness of a hypothetical local

TMDC layer that reproduces the same polariton wavevector in an otherwise local air/TMDC($d$)/α-MoO$_3$($t$)/SiO$_2$ stack (Supplementary Fig. 1). After extracting the nonlocal polariton wavevector $q_{NL}(t)$ from the Feibelman $d_\perp$ framework, inverting the local multilayer dispersion yields the analytic mapping

$$t_{\text{eff}}(t) = \frac{i}{2k_{z1}(q_{NL}(t))} \ln\left[\frac{r_{21}(q_{NL}(t)) - r_{2d}^*(t)}{r_{10}(q_{NL}(t))(r_{2d}^*(t)r_{21}(q_{NL}(t)) - 1))}\right] \quad (1)$$

where $k_{z1}$ is the out-of-plane wavevector component in the TMDC and $r_{ij}$ are Fresnel reflection coefficients evaluated at $q_{NL}(t)$ (full definitions in Methods and Supplementary Section 1). The resulting $t_{\text{eff}}(t)$ reports, within a local-response language, how much of a physical 25-nm WSe$_2$ flake effectively participates in polariton confinement as the α-MoO$_3$ thickness is tuned (Extended Data Fig. 3). Consistent with the $d_\perp$ analysis, $t_{\text{eff}}$ remains close to the physical thickness in the depletion regime but drops sharply upon entering the saturation plateau, providing an independent local-response consistency check.

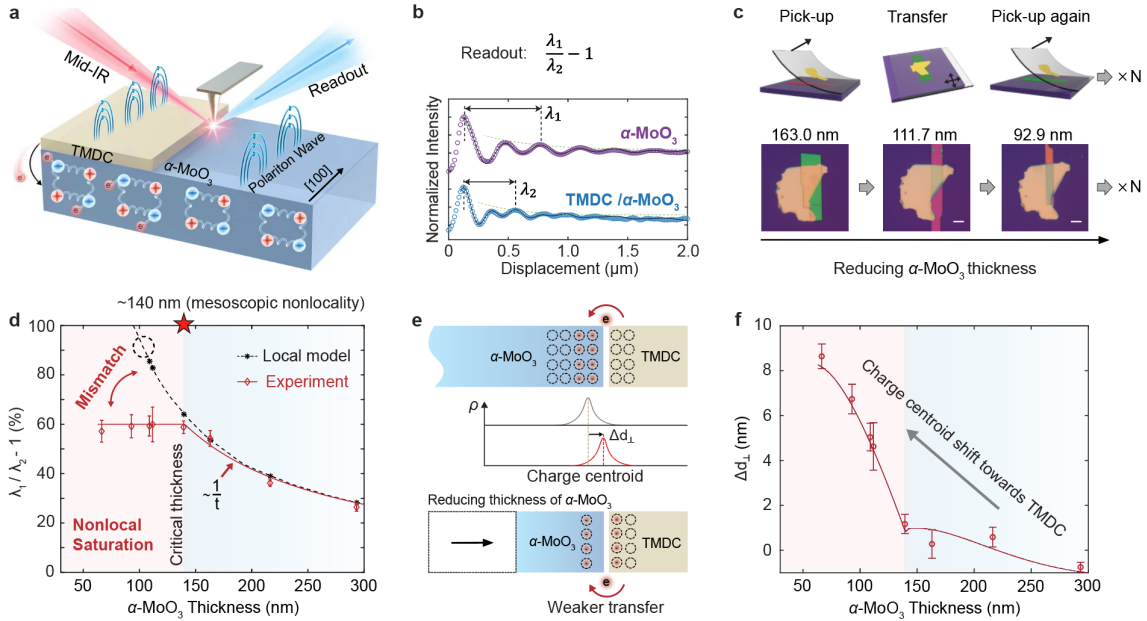

**Fig. 1. Mesoscopic nonlocal screening and thickness-independent saturation in charge-transfer polaritonic heterostructures. a**, Polaritons as near-field probes of buried TMDC/α-MoO$_3$ interfaces, interfacial charge transfer and charge redistribution in α-MoO$_3$ renormalize surface phonon polaritons (PhPs), which are directly imaged by scattering-type scanning near-field optical microscopy (s-SNOM). **b**, Definition of the polaritonic readout, $\lambda_1/\lambda_2 - 1$, extracted from representative s-SNOM line profiles; $\lambda_1$ and $\lambda_2$ denote the PhPs wavelengths on bare α-MoO$_3$ and under TMDC coverage, respectively. **c**, Capillary-force-assisted reconfigurable transfer workflow enabling the same TMDC flake to be sequentially positioned on multiple α-MoO$_3$ flakes with different thicknesses (examples shown). Scale bars, 10 μm. **d**, Thickness dependence of the

readout for WSe$_2$/α-MoO$_3$. A depletion-like $1/t$ trend at large α-MoO$_3$ thickness transitions to a thickness-independent plateau below a critical thickness of ~140 nm, revealing a mesoscopic nonlocal regime far beyond canonical ångström-nanometre nonlocal corrections and not captured by a local fixed-permittivity model (dashed curve). **e**, Feibelman $d_\perp$ picture of screening saturation: thinning α-MoO$_3$ broadens the induced charge distribution and shifts its centroid towards the TMDC interface, quantified by $\Delta d_\perp$. **f**, The extracted $\Delta d_\perp$ as a function of α-MoO$_3$ thickness, showing a pronounced increase upon entering the saturation regime, consistent with nonlocal charge redistribution underpinning the polariton plateau.

**Establishing a universal polariton probe**

Rather than merely complicating predictive optical modelling, the nonlocal saturation turns a potential limitation into an opportunity. When the polaritonic host α-MoO$_3$ reaches its intrinsic saturation regime, the near-field readout becomes independent of α-MoO$_3$ thickness and therefore directly comparable across different interfaces. This provides a practical route for benchmarking buried charge-transfer interfaces on equal footing, without ambiguity from partially screened, thickness-dependent host responses. To establish such a cross-material universal polariton probe, we use three thin α-MoO$_3$ hosts within the saturation regime to identify, for each TMDC, an α-MoO$_3$-thickness-independent saturated readout (Fig. 2a).

We then apply this saturated readout across a series of TMDC/α-MoO$_3$ interfaces and observe a systematic trend. For TMDC flakes of comparable thickness (approximately 9 nm), the saturated polariton wavelength decreases from MoS$_2$ to MoSe$_2$ and to MoTe$_2$, consistent with progressively stronger interfacial charge transfer (Fig. 2c). The same trend holds across the full set of five MX$_2$ (M = Mo or W; X = S, Se, or Te), including WS$_2$ and WSe$_2$, which are shown in Extended Data Fig. 2, and Supplementary section 5. To connect the saturated readout to an interfacial driving force, we consider the work-function difference $\Delta\Phi$ between the TMDC and α-MoO$_3$, which sets the built-in electrostatic bias for charge redistribution under a vacuum-level alignment picture (Fig. 2b). The saturated readout increases with $\Delta\Phi$ and is approximately linear across the materials studied (Fig. 2d), indicating that larger work function difference drives stronger interfacial polarization and hence a larger modulation of the PhP wavelength. Together, this thickness-invariant calibration establishes the saturated polariton probe as a quantitative optical ruler for comparing buried vdW interfaces across disparate heterostructures.

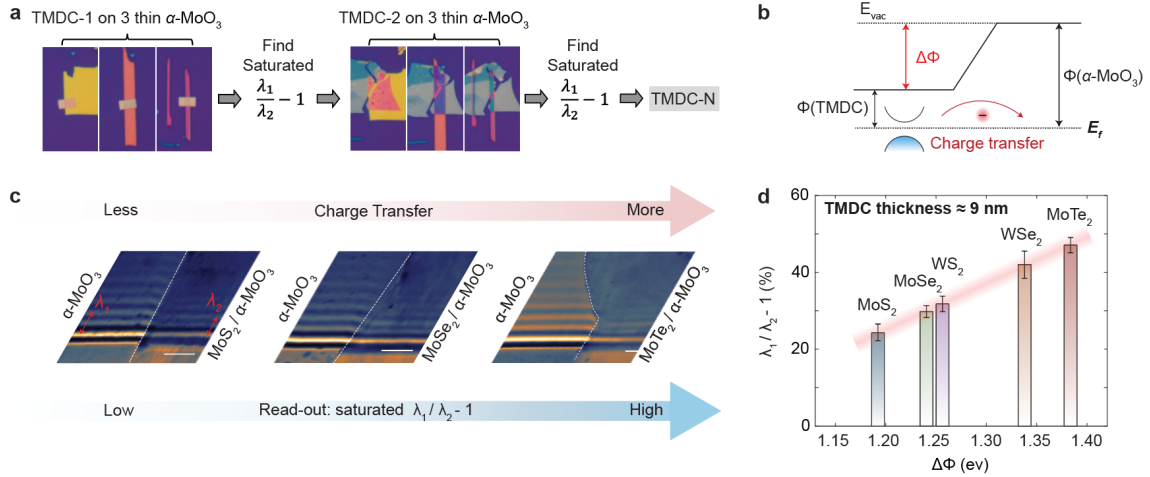

**Fig. 2. A saturated polariton readout enables cross-material quantification of interfacial charge transfer. a**, Workflow for extracting a host-thickness-independent (saturated) readout for each TMDC, the readout $\lambda_1/\lambda_2 - 1$ is measured on three thin $\alpha$-MoO$_3$ hosts within the saturation regime to verify convergence and obtain a comparable saturated value. **b,** Work-function difference $\Delta\Phi$ between a TMDC and $\alpha$-MoO$_3$ sets the built-in driving force for electron transfer upon contact under a vacuum-level alignment picture. **c,** Representative s-SNOM images of PhPs across MoS$_2$/$\alpha$-MoO$_3$, MoSe$_2$/$\alpha$-MoO$_3$ and MoTe$_2$/$\alpha$-MoO$_3$ interfaces, showing progressively stronger wavelength reduction under TMDC coverage ($\lambda_2$) relative to bare $\alpha$-MoO$_3$ ($\lambda_1$) with increasing charge transfer. Scale bars, 1 μm. **d,** Saturated readout versus $\Delta\Phi$ for ~9 nm-thick MX$_2$ (M = Mo or W; X = S, Se, or Te) overlayers, revealing an approximately linear scaling that calibrates PhPs as an optical ruler for buried interfacial electrostatics

**Cross-material validation of the universal polariton probe**

To test the reliability of the saturated polariton probe across materials and thicknesses, we built a device library by sequentially transferring multiple semiconductor flake onto $\alpha$-MoO$_3$ host of systematically varied thickness. We fabricated more than 120 heterostructures by transferring six WSe$_2$ and seven PdSe$_2$ flakes onto eight $\alpha$-MoO$_3$ substrates. WSe$_2$ serves as a representative TMDC whose work function varies with thickness, whereas PdSe$_2$ provides a structurally distinct, orthorhombic material platform that undergoes a thickness-driven semiconductor-to-semimetal transition beyond ~30 nm[44]. Representative transfer sequences for a 20.4 nm-thick WSe$_2$ and a 27.9 nm PdSe$_2$ flake are shown in Fig. 3a, e, and the full raw dataset is provided in Supplementary Figs. 13-31.

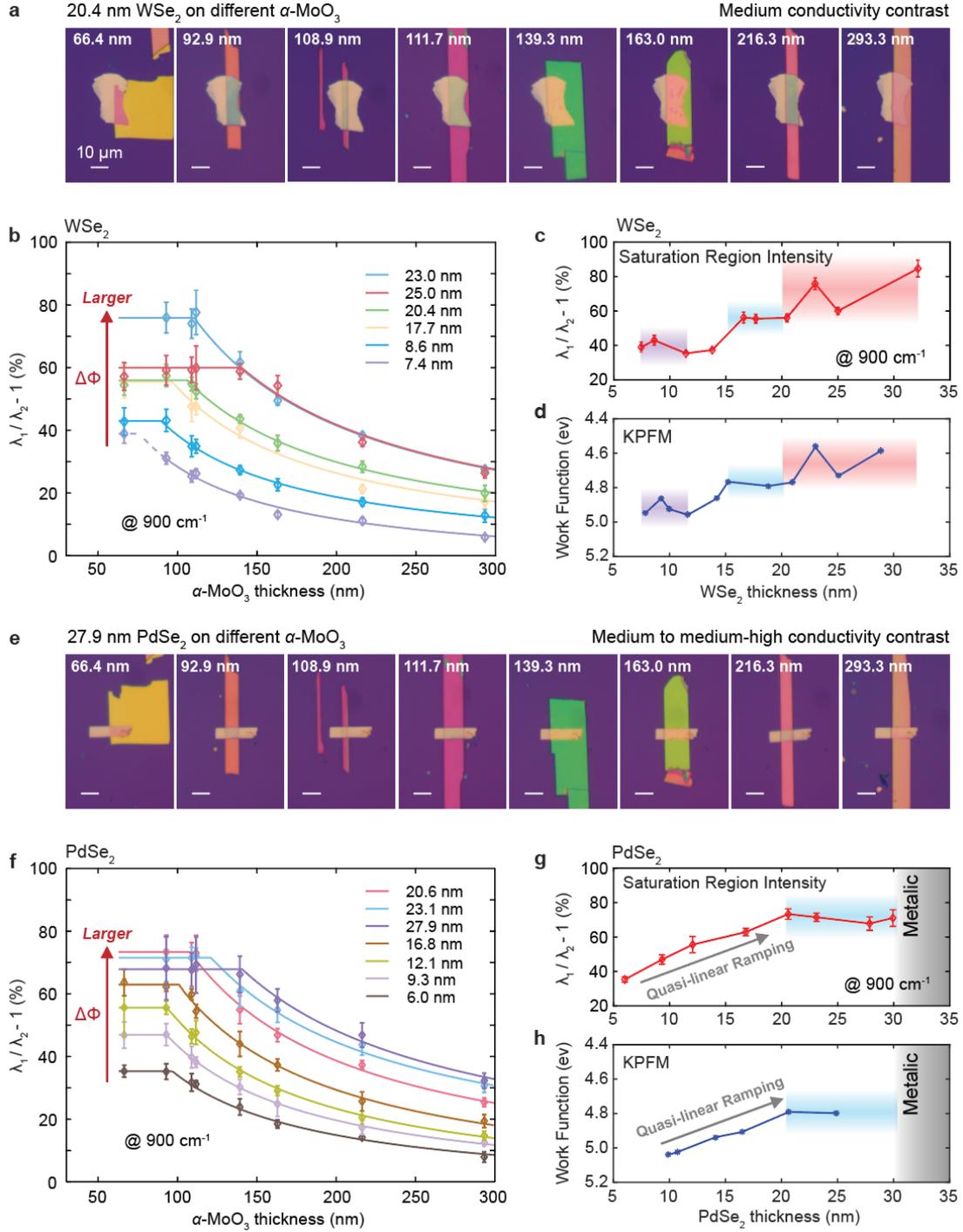

Fig. 3. Systematic investigation of the saturation and depletion regions spans hexagonal and orthorhombic lattice TMDC/α-MoO₃ heterostructures, respectively. a, e, Optical images of representative heterostructures: medium conductivity contrast (WSe₂/α-MoO₃) and medium-to-high conductivity contrast (PdSe₂/α-MoO₃), TMDC flakes (~20-30 nm thickness) transferred onto α-MoO₃ substrates with varying thicknesses. Scale bars: 10 μm. b, f, Polaritonic readout ($\lambda_1/\lambda_2 - 1$) versus α-MoO₃

thickness for multiple WSe$_2$ and PdSe$_2$ thicknesses, clearly exhibiting distinct saturation and depletion regions. **c, g,** Saturated readout for WSe$_2$ and PdSe$_2$ as a function of TMDC thickness. **d, h,** Corresponding TMDC work functions measured by Kelvin Probe Force Microscopy (KPFM). Saturated readout correlates strongly with TMDC work functions.

Fig. 3b summarizes the polaritonic readout $\lambda_1/\lambda_2 - 1$ for six WSe$_2$ device sets formed by combining a fixed WSe$_2$ flake (23.0, 25.0, 20.4, 17.7, 8.6, and 7.4 nm) with the same panel of eight α-MoO$_3$ hosts. Across all sets, we consistently observe a depletion-like branch at large α-MoO$_3$ thickness and a saturation plateau below flake-dependent critical thickness. This reproducibility shows that the saturation-based readout is robust across repeated transfers and across devices.

Comparisons between WSe$_2$ flakes with similar thickness but different work function isolate the role of the built-in driving force $\Delta\Phi$. For example, the depletion branches of the 23.0- and 25.0-nm WSe$_2$ sets largely overlap, whereas their saturation plateaus separate when α-MoO3 becomes thinner than ~140 nm. Kelvin Probe Force Microscopy (KPFM) shows that the 23.0-nm WSe2 flake has a lower work function than the 25.0-nm flake (Supplementary Figs. 10-12), corresponding to a larger $\Delta\Phi$ against α-MoO$_3$ and therefore a higher saturated readout. By contrast, the 20.4- and 17.7-nm WSe2 sets have similar work functions and correspondingly similar saturated plateaus, while the thicker flake still exhibits a larger critical thickness.

To refine the $\Delta\Phi$ dependence specifically within the saturation regime, we additionally measured four WSe$_2$ flakes (11.4, 13.8, 16.6, and 32.2 nm) transferred only onto the three thinnest α-MoO$_3$ hosts, thereby directly sampling the saturated readout without contributions from the depletion branch. Figs. 3c, d compile the saturated readout and corresponding work function as a function of WSe2 thickness, revealing consistent thickness windows and a pronounced maximum around ~23 nm.

We next test the generality of these behaviours in PdSe$_2$, an orthorhombic layered material that contrasts sharply with hexagonal TMDCs and forms orthorhombic-to-orthorhombic interfaces with α-MoO$_3$[44,45]. We fabricated seven PdSe$_2$/α-MoO$_3$ devices sets (PdSe$_2$ thickness 20.6, 23.1, 27.9, 16.8, 12.1, 9.3, and 6.0 nm) and again observe the same two-regime behaviour with a thickness-independent saturation plateau and a thickness-dependent crossover. Notably, PdSe$_2$ flakes in the 20-30 nm range exhibit a quasi-constant work function, and the 20.6-, 23.1- and 27.9-nm sets therefore show similar saturated readouts while the thicker flake exhibits a larger critical thickness (Fig. 3f). This agreement demonstrates that saturation-based benchmarking extends beyond hexagonal TMDCs to materials with distinct lattice symmetry.

**Revising Anderson's rule for vdW heterostructures**

By analyzing the full device library, we found that the saturated polaritonic metric $\lambda_1/\lambda_2 - 1$ correlates linearly with the work-function difference $\Delta\Phi$ for both 2H-MX$_2$/α-MoO$_3$ and PdSe$_2$/α-MoO$_3$ heterostructures. In Fig. 4a, the two material families show comparable slopes but distinct horizontal intercepts, revealing different onset energies for appreciable interfacial charge transfer. Linear extrapolation of each family to $\lambda_1/\lambda_2 - 1 = 0$ yields an onset $\Delta\Phi_0$, and the data are well described by the empirical relation:

$$\frac{\lambda_1}{\lambda_2} - 1 (\text{saturated}) = 137(\%) * \left(\Phi_{\alpha-\text{MoO}_3} - \Phi_{\text{TMDC}}\right) - \Delta\Phi_0 \qquad (2)$$

where $\Phi_{TMDC}$ and $\Phi_{\alpha-\text{MoO}_3}$ denote the work functions of TMDC and α-MoO$_3$ respectively, S is the fitted slope, and $\Delta\Phi_0$ is a lattice-mismatch-dependent onset energy. Physically, $\Delta\Phi_0$ defines the barrier-like onset that must be overcome before measurable interfacial charge transfer occurs. We obtain $\Delta\Phi_0$ = 0.72 eV for PdSe2 and $\Delta\Phi_0$ = 0.92 eV for hexagonal MX$_2$. Notably, across more than 120 devices spanning different twist angles and lateral registries, the extracted onset varies little within each family, suggesting that it is set primarily by interfacial lattice compatibility rather than stacking geometry within the explored range.

We associate the difference $\Delta\Phi_0$ with the degree of lattice compatibility at the buried interface. As shown in Fig. 4d, e, α-MoO$_3$ and PdSe$_2$ are both orthorhombic but have different lattice constants, whereas 2H-MX$_2$ (M = Mo or W; X = S, Se, or Te) is hexagonal. A near-commensurate coincidence exists for PdSe$_2$/α-MoO$_3$, for which 2a(PdSe$_2$) is close to 3a(α-MoO$_3$), while no analogous simple commensurability is apparent for 2H-MX$_2$/α-MoO$_3$. This structural contrast provides a natural explanation for the smaller onset in the more lattice-compatible interface.

These observations motivate a revision of Anderson's rule (Fig. 4b) for vdW interfaces. The traditional vacuum-level alignment picture, often referred to as Anderson-type band alignment, was developed for bulk, lattice-matched junctions in which a separate transfer onset is typically negligible[46,47]. In contrast, the lattice-mismatch-tolerant assembly of vdW heterostructures permits arbitrary lattice combinations, and our statistics reveal a barrier-like onset $\Delta\Phi_0$ that depends on interfacial lattice compatibility. We therefore define an effective driving force $\Delta\Phi_{\text{eff}} = \Delta\Phi - \Delta\Phi_0$ and use it to revise Anderson-type alignment for vdW interfaces (Fig. 4c), yielding improved predictive power for charge-transfer-driven interface modulation.

Finally, the same global analysis links the local-to-saturation crossover (the critical α-MoO3 thickness) to both TMDC thickness, and the effective driving force $\Delta\Phi_{\text{eff}}$, supporting a unified picture in which vdW-specific lattice terms and dimensional screening jointly govern buried charge-transfer behaviour.

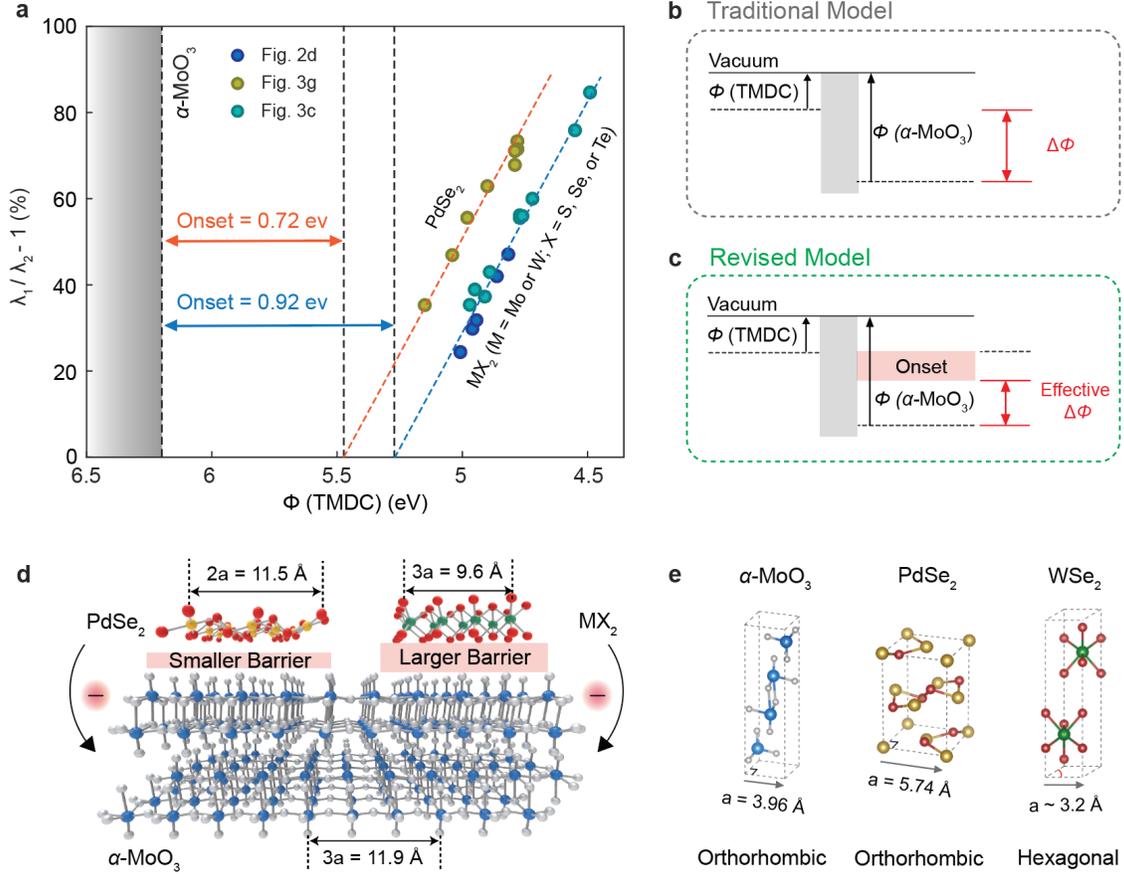

**Fig. 4. Lattice-mismatch-induced barrier-like onset for interfacial charge transfer.** **a**, Correlation between the saturated polaritonic readout $\lambda_1/\lambda_2 - 1$ and the work-function difference $\Delta\Phi$, revealing two family-dependent onset energies $\Delta\Phi_0$ for appreciable charge-transfer-driven modulation (~0.72 eV for $\alpha$-MoO$_3$/PdSe$_2$ and ~0.92 eV for 2H-MX$_2$/$\alpha$-MoO$_3$). **b**, Conventional Anderson-type vacuum-level alignment, in which $\Delta\Phi$ directly sets the driving force for charge transfer. **c**, Revised vdW-specific picture incorporating an onset $\Delta\Phi_0$ and an effective driving force $\Delta\Phi_{\text{eff}} = \Delta\Phi - \Delta\Phi_0$. **d**, Atomic-scale schematic highlighting contrasting interfacial lattice compatibility. **e**, Crystal structures of $\alpha$-MoO$_3$/PdSe$_2$, PdSe$_2$ and WSe$_2$, illustrating the in-plane lattice relationships that underlie the different onset energies.

## Conclusion

By using polaritons as quantitative probes of buried charge-transfer interfaces, we reveal a mesoscopic nonlocal electrostatic regime that directly challenges the conventional view of nonlocality as ångström-nanometre-scale corrections. In TMDC/$\alpha$-MoO$_3$ heterostructures, this regime appears as a thickness-independent saturation plateau below a critical $\alpha$-MoO$_3$ thickness of up to ~140 nm and can be described within a Feibelman $d$-parameter framework. Once the host reaches this intrinsic saturation state, the polariton readout becomes a universal optical metric for buried vdW interfaces. Across more than 120 devices, this metric shows an approximately linear dependence on work-function difference and reveals a lattice-mismatch-dependent onset energy for charge transfer.

More broadly, this polaritonic readout and the revised vdW-Anderson framework provide a practical basis for treating band alignment, charge transfer and screening as quantitative design parameters in vdW nanodevices. This approach enables interface-aware interpretation of contacts in 2D electronics and optoelectronics, constrains interface-induced dielectric renormalization in polaritonic and meta-optical platforms, and offers transferable design rules for the rapidly expanding vdW heterostructure landscape.

**Methods**

**Fabrication of heterostructure devices**

TMDC flakes and multilayered $\alpha$-MoO$_3$ flakes were mechanically exfoliated from bulk crystals onto 295 nm thick SiO$_2$/Si wafers. The 2H-MoS$_2$, 2H-MoSe$_2$, 2H-WS$_2$, 2H-WSe$_2$, and 2H-MoTe$_2$ crystals were purchased from 2D Semiconductors, Inc. Bulk orthorhombic PdSe$_2$ single crystals were synthesized via the self-flux method in a horizontal furnace maintained at 850 °C for seven days. $\alpha$-MoO$_3$ single crystals were grown employing the physical vapor transport method in a two-zone furnace with temperature settings of 750 °C and 650 °C. All flakes were identified under an optical microscope for approximate thickness and verified using atomic force microscopy (AFM). The AFM images are provided in Supplementary Figs. 6, 8 and 9. All flakes were transferred using the refined capillary-force-assisted clean-stamp transfer method based on our previous work[39,48]. For the reconfigurable transfer, we meticulously control the water vapor, so that the TMDC flake can be selectively picked up while the $\alpha$-MoO$_3$ remains on the substrate. This selectivity is due to the $\alpha$-MoO$_3$ flake having much lower contact angles compared with those of TMDCs.

**Scattering-type scanning near-field optical microscopy (s-SNOM) measurements**

Near-field imaging for extracting the readout $\lambda_1/\lambda_2 - 1$ was performed using a commercially available s-SNOM (NeaSpec) equipped with a mid-infrared quantum cascade laser (Daylight Solutions) under ambient conditions[24]. The metal-coated AFM probe was used to enhance the incident p-polarized mid-IR light, which acts as an antenna for signal collecting. The TMDC/$\alpha$-MoO$_3$ heterostructure samples were raster-scanned below the oscillating tip at a tapping frequency of ~270 kHz and an amplitude of ~65 nm. The tip-scattered signal was then recorded by thermal detectors and demodulated at the second and third harmonic of the tapping frequency (s2 and s3 signal channels, respectively) to minimise the far-field background noise.

**Kelvin Probe Force Microscopy (KPFM) measurements**

KPFM measurements were conducted using an atomic force microscope (AFM) operating in electrostatic force microscopy (EFM) mode (Asylum Research Cypher S). Conductive tips coated with a Pt/Ir film, featuring a resonance frequency of ~70 kHz and a force constant of ~2 N/m, were employed. The first step was to transfer exfoliated TMDC flakes onto an Au electrode. Then, the topography of the exfoliated sample was collected under contact mode. Subsequently, the AFM mode was switched to EFM mode to record the potential difference between the tip and the samples, with the Au film (work function 5.1 eV[49]) and 60 nm thick graphite flake (work function 4.7 eV[50]) serving as the reference standards.

**Numerical simulations**

In this study, we employed the finite-element method (FEM) using COMSOL Multiphysics (version 6.0) to simulate near-field images and calculate the readout $\lambda_1/\lambda_2 - 1$. We modelled the TMDC/$\alpha$-MoO$_3$ heterostructures as a pair of dielectric layers, each with specific thickness values. These thickness values were determined via atomic force microscopy (AFM). The permittivity tensor for $\alpha$-MoO$_3$ was initially based on literature values and subsequently adjusted in accordance with our near-field measurements[51]. This refinement was essential to ensure that the permittivity tensor accurately reflected the specific conditions of our crystals, as opposed to those detailed in the reference. The in-plane and out-of-plane permittivity values of WSe$_2$ were set at 12.0 and 5.8, respectively. These were calibrated to align with the experimental $\lambda_1/\lambda_2 - 1$ ratio observed in the 25.0-nm-WSe$_2$/139.3-nm-$\alpha$-MoO$_3$ device (Fig. 2b) and were based on DFT calculations. It is noteworthy that current methodologies do not provide precise permittivity values for thick TMDCs. The primary goal of our numerical simulations of $\lambda_1/\lambda_2 - 1$ was to demonstrate the trend of $\lambda_1/\lambda_2 - 1 \sim 1/t$ ($t$ denotes the thickness of the $\alpha$-MoO$_3$) in the trivial depletion region, acknowledging that this model is invalid in the non-trivial, saturated region.

**Feibelman *d*-parameter framework for charge screening saturation**

To capture the breakdown of local-response screening and the thickness-independent saturation regime, we model the TMDC/$\alpha$-MoO$_3$ interface using the Feibelman *d*-parameter (surface-response) formalism[41,43]. The key quantity is the out-of-plane parameter $d_\perp$, interpreted as the centroid of the induced charge density relative to the geometrical interface plane[42].

We consider a three-layer TM-mode geometry (Region I: air; Region II: $\alpha$-MoO$_3$ slab of thickness t; Region III: SiO$_2$). In the local-response limit, applying Maxwell boundary conditions at the two interfaces yields the transfer-matrix dispersion relation:

$$e^{-i2k_{z2}t} = \frac{\left(\frac{k_{z2}}{\varepsilon_{x,II}}-k_{z1}\right)\left(\frac{k_{z2}}{\varepsilon_{x,II}}-\frac{k_{z3}}{\varepsilon_{III}}\right)}{\left(\frac{k_{z2}}{\varepsilon_{x,II}}+k_{z1}\right)\left(\frac{k_{z2}}{\varepsilon_{x,II}}+\frac{k_{z3}}{\varepsilon_{III}}\right)} \quad (3)$$

Including mesoscopic nonlocality modifies the interface conditions through Feibelman-like surface-response lengths $d_\perp$ and $d_\parallel$ leading to the generalized nonlocal dispersion relation:

$$e^{-i2k_{z2}t} = \frac{\left[\beta_2\left(-\frac{k_{z3}}{\varepsilon_{III}}-d_\perp\frac{ik_x^2}{\varepsilon_{III}}\right)-\left(-\frac{k_{z2}}{\varepsilon_{x,II}}-d_\perp\frac{ik_x^2}{\varepsilon_{z,II}}\right)\right]\left[(k_{z1}-id_\perp k_x^2)\alpha_1-\left(\frac{k_{z2}}{\varepsilon_{x,II}}-d_\perp\frac{ik_x^2}{\varepsilon_{z,II}}\right)\right]}{\left[\alpha_2\left(-\frac{k_{z3}}{\varepsilon_{III}}-d_\perp\frac{ik_x^2}{\varepsilon_{III}}\right)-\left(\frac{k_{z2}}{\varepsilon_{x,II}}-d_\perp\frac{ik_x^2}{\varepsilon_{z,II}}\right)\right]\left[(k_{z1}-id_\perp k_x^2)\beta_1-\left(-\frac{k_{z2}}{\varepsilon_{x,II}}-d_\perp\frac{ik_x^2}{\varepsilon_{z,II}}\right)\right]} \quad (4)$$

with the auxiliary factors:

$$\alpha_2 = \frac{(1+id_\parallel k_{z2})}{(1-id_\parallel k_{z3})}, \beta_2 = \frac{(1-id_\parallel k_{z2})}{(1-id_\parallel k_{z3})}, \alpha_1 = \frac{(1+id_\parallel k_{z2})}{(1+id_\parallel k_{z1})}, \beta_1 = \frac{(1-id_\parallel k_{z2})}{(1+id_\parallel k_{z1})} \quad (5)$$

In the analysis reported here we neglect in-plane nonlocality by setting $d_\parallel = 0$, and retain only $d_\perp$ as the nonlocal length scale.

For each $\alpha$-MoO$_3$ thickness, we solve the dispersion relation at the experimental frequency (900 cm$^{-1}$) to obtain the in-plane polariton wavevector $q = kx$ and thus the wavelength $\lambda = \frac{2\pi}{Re(q)}$. The polaritonic readout is computed as $\lambda_1/\lambda_2 - 1$, where $\lambda_1$ and $\lambda_2$ are the wavelengths on bare $\alpha$-MoO$_3$ and under TMDC coverage, respectively (Fig. 1b). We extract an effective $d_\perp$ by fitting the measured $\lambda_1/\lambda_2 - 1$ with fixed bulk permittivities and measured layer thicknesses; $d_\perp$ is the only free parameter.

As a local-response consistency check (Extended Data Fig. 3), we additionally fit the same $\lambda_1/\lambda_2 - 1$ data within a fixed-permittivity layer-stacking model by allowing an effective TMDC thickness to vary as a phenomenological fitting parameter. For each $\alpha$-MoO$_3$ thickness, $d_\perp$ is extracted by fitting the measured $\lambda_1/\lambda_2 - 1$ while keeping all geometrical layer thicknesses fixed to AFM values and all bulk permittivities fixed to the values used in the local-response model.

**First-principles calculations**

The charge density difference and Bader charge analysis among heterostructures: monolayer graphene/ four-layer $\alpha$-MoO$_3$ and MX$_2$/$\alpha$-MoO$_3$ are obtained with spin-polarized density functional theory (DFT) calculations[52]. The monolayer graphene/four-layer $\alpha$-MoO$_3$ heterostructures are built by $\sqrt{2} \times \sqrt{2} \times 2$ MoO$_3$ supercell and $2\sqrt{2} \times \sqrt{2} \times 1$ graphene supercell with a 1.5nm vacuum slab.

Other heterostructures are built by $\sqrt{2} \times \sqrt{2} \times 2$ $\alpha$-MoO$_3$ supercell and $2\sqrt{2} \times \sqrt{2} \times 1$ MX$_2$ supercell with 1.5nm vacuum slab.

DFT calculations were conducted using the Vienna ab initio Simulation Package (VASP)[53] with on-site Coulomb interaction (DFT + U) corrected for Mo 4d and W 5d electrons[54]. $U_{eff}$ = 4.39 eV and $U_{eff}$ = 6.3 eV were applied to Mo 4d and W 5d orbitals, respectively. The projector-augmented wave (PAW) potentials[55] were used, and the exchange-correlation part of the Kohn-Sham equation was treated at the generalized gradient approximation (GGA) level using the Perdew-Burke-Ernzerhof (PBE) functional[56]. The Grimme D3 scheme (DFT-D3[57]) was employed to correct van der Waals (vdW) interactions. The kinetic energy cutoff for the plane wave basis set was set to 500 eV. For electronic minimization, total energy was converged within $10^{-5}$ eV and electron smearing was treated using the Gaussian smearing method with a smearing width of 0.05 eV. The Monkhorst-Pack $k$-point meshes with a k-point spacing of 0.04 Å$^{-1}$ were applied for the Brillouin-zone integration. Structures were relaxed using the conjugate gradient method until their interatomic forces were less than 0.02 eV/Å.

**Data availability**

All data that support the findings of this study are available in the main text, figures, and Supplementary Information. They are also available from the corresponding author upon reasonable request.

**Code availability**

All code that are used in the main text and Supplementary Information are available from the corresponding author upon reasonable request.

**Author contributions**

X.M. and Q.W. conceived the idea; Q.W., Z. L. and Y. L. and X. M. supervised the project; X.M. developed the theory and models; X.M. carried out the numerical simulations with assistance from Y.Z.; X.M. developed the reconfigurable clean stamp transfer method and fabricated the heterostructure devices; R.D. synthesized the $\alpha$-MoO$_3$ and PdSe$_2$ crystals. Z.L. and X.M. carried out the s-SNOM characterization; X.M. analysed the data with the help of Q.L. M.J. and Y.M.; Z.D. and Z.L. carried out some parts of DFT simulations. Huasuan Company carried out the rest of the DFT simulation under supervision from X.M. and M.J.; X.M. and M.J. prepared the KPFM samples. J.S. and R.D. carried out the KPFM scanning with help from X.M. M.J and Z.L. X.M. wrote the manuscript with input from Z. L and Z.D. All authors discussed the results and the manuscript.

**Competing Interests**

The authors declare no competing interests.

**Acknowledgments**

We acknowledge the Agency for Science, Technology and Research (ASTAR) for providing core funding to support the facilities. **Funding**: AME IRG Grant No. M22k2c0080, MTC Programmatic Project Grant No. M23M2b0056, Career Development Fund (CDF) Seed Grant No. 222D800038, and MTC Young Individual Research Grant (YIRG) Nos. M23M7c0129, M24N8c0097, and M23M7c0119; MTC Individual Research Grant (IRG) Nos. M24N7c0081 and M24N7c0087. A*STAR Quantum Innovation Centre (Q. InC) core funding; Singapore National Research Foundation (NRF) Competitive Research Program (CRP) Grants NRF-CRP22-2019-0007, NRF-CRP31-0001, and NRF-CRP264-2021-0004; Singapore Ministry of Education (MOE) Tier 3 Programmatic Fund Award MOE-MOET32023-0003 and Tier 2 Grant MOE-T2EP50223-0008, MOE-T2EP50224-0044, and MOE-T2EP50125-0038; Singapore MOE Academic Research Fund (AcRF) Tier 1 Award No. 023785-00001; National Natural Science Foundation of China Grant No. 12404363; Natural Science Foundation of Jiangsu Province Grant No. BK20241374; and Fundamental Research Funds for the Central Universities (Nanjing University of Aeronautics and Astronautics) Grant Nos. NE2024007 and NS2024022.

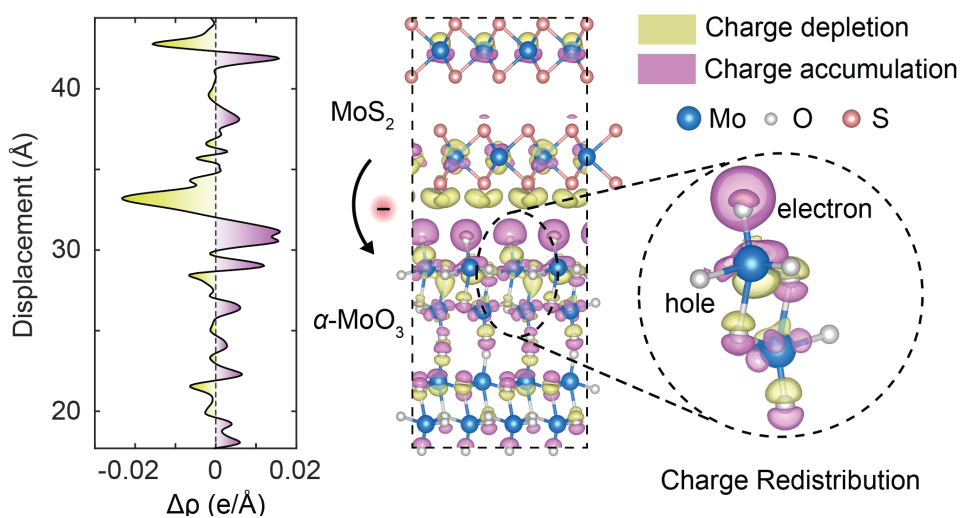

**Extended Data Fig. 1. DFT-calculated charge redistribution with extended length in the MoS$_2$/$α$-MoO$_3$ heterostructure.** The left panel shows the plane-averaged charge density difference $Δρ(z)$ projected along the stacking axis, with regions of electron depletion (yellow) and accumulation (purple). The centre panel

displays a three-dimensional Δρ isosurface, revealing the formation of a dipole across the interface. The right inset highlights the interfacial charge transfer, with electrons accumulating on the α-MoO$_3$ side and holes on the MoS$_2$ side; Mo, O and S atoms are coloured blue, white and orange, respectively. This result illustrates the formation of an interface dipole consistent with the interfacial polarization observed in the main text and supports the proposed physical mechanism underlying the polariton wavelength shift and asymmetric screening.

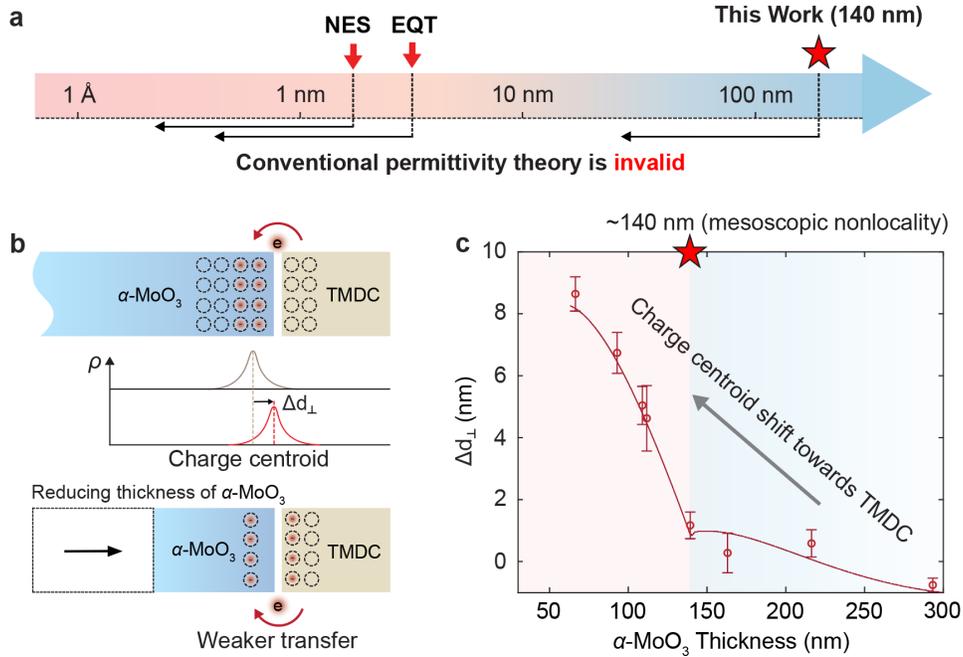

**Extended Data Fig. 2. Mesoscopic nonlocal screening length scale from Feibelman charge-centroid analysis. a**, Comparison of the ~140 nm mesoscopic nonlocality identified here with canonical ångström-nanometre nonlocal effects in extreme plasmonics (NES, nonlocal electronic screening; EQT, electron quantum tunnelling), highlighting that a conventional local, layer-by-layer permittivity description can break down over much larger length scales in these phonon-polaritonic vdW heterostructures. b, Schematic of charge transfer at a TMDC/α-MoO$_3$ buried interface and the associated displacement $\Delta d_\perp$ of the induced-charge centroid from the geometrical interface plane. Reducing the α-MoO$_3$ thickness modifies the screening volume and shifts the induced-charge centroid towards the TMDC side. **c**, Extracted $\Delta d_\perp$ as a function of α-MoO$_3$ thickness, showing a crossover at ~140 nm into a mesoscopic nonlocal regime. Shaded regions indicate the saturation regime (pink) and depletion regime (blue). Error bars denote standard deviations from repeated extractions.

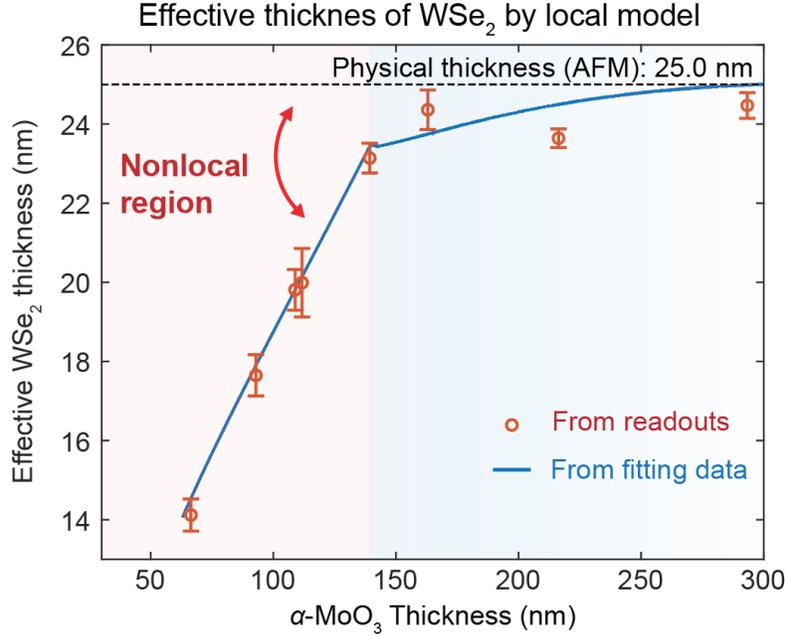

**Extended Data Fig. 3. Effective thickness of WSe2 extracted from the local model.** Effective thickness is obtained from the near-field polariton readout ($\lambda_1/\lambda_2 - 1$), using a local dielectric description and is plotted as a function of the underlying α-MoO₃ thickness. It quantifies the portion of WSe₂ that effectively participates in interfacial polarization. A clear saturation emerges at thin α-MoO₃ t thicknesses (pink shaded region), indicating a marked reduction of charge transfer. For thicker α-MoO₃ (blue shaded region), the effective thickness progressively approaches the physical 25 nm thickness of the WSe₂ flake, exhibiting conventional depletion-like behaviour (~$1/t$ dependence). The transition point (up to ~140 nm of α-MoO₃) , matches the saturation threshold observed from the nonlocal screening analysis, confirming a nonlocal mechanism for interfacial polarization saturation. Error bars represent standard deviations from multiple measurements.

**Reference**

1	Geim, A. K. & Grigorieva, I. V. Van der Waals heterostructures. *Nature* **499,** 419-425 (2013).
2	Liu, Y. *et al*. Van der Waals heterostructures and devices. *Nat. Rev. Mater.* **1,** 16042 (2016).
3	Novoselov, K. S., Mishchenko, A., Carvalho, A. & Castro Neto, A. H. 2D materials and van der Waals heterostructures. *Science* **353,** aac9439 (2016).
4	Rivera, P. *et al*. Valley-polarized exciton dynamics in a 2D semiconductor heterostructure. *Science* **351,** 688-691 (2016).
5	Cao, Y. *et al*. Unconventional superconductivity in magic-angle graphene superlattices. Nature 556, 43-50 (2018).
6	Li, W. *et al.* Approaching the quantum limit in two-dimensional semiconductor contacts. *Nature* **613,** 274-279 (2023).
7	Basov, D. N., Fogler, M. M. & García de Abajo, F. J. Polaritons in van der Waals materials. *Science* **354,** aag1992 (2016).
8	Du, L. *et al.* Moiré photonics and optoelectronics. *Science* **379,** eadg0014 (2023).

9   Sternbach, A. J. et al. Negative refraction in hyperbolic hetero-bicrystals. *Science* **379,** 555-557 (2023).
10  Hu, H. *et al.* Gate-tunable negative refraction of mid-infrared polaritons. *Science* **379,** 558-561 (2023).
11  Dai, S. *et al.* Tunable Phonon Polaritons in Atomically Thin van der Waals Crystals of Boron Nitride. *Science* **343,** 1125-1129 (2014).
12  Bernardi, M., Ataca, C., Palummo, M. & Grossman, J. C. Optical and Electronic Properties of Two-Dimensional Layered Materials. *Nanophotonics* **6,** 479-493 (2017).
13  Kim, B. S. Y. *et al.* Ambipolar charge-transfer graphene plasmonic cavities. *Nat. Mater.* **22,** 838–843 (2023).
14  Babaze, A., Neuman, T., Esteban, R., Aizpurua, J. & Borisov, A. G. Dispersive surface-response formalism to address nonlocality in extreme plasmonic field confinement. *Nanophotonics* **12,** 3277-3289 (2023).
15  Luo, Y., Fernandez-Dominguez, A. I., Wiener, A., Maier, S. A. & Pendry, J. B. Surface Plasmons and Nonlocality: A Simple Model. *Phys. Rev. Lett.* **111,** 093901 (2013).
16  Savage, K. J. *et al.* Revealing the quantum regime in tunnelling plasmonics. *Nature* **491,** 574-577 (2012).
17  Ciracì, C. *et al.* Probing the Ultimate Limits of Plasmonic Enhancement. *Science* **337,** 1072-1074 (2012).
18  Lundeberg, M. B. *et al.* Tuning quantum nonlocal effects in graphene plasmonics. *Science* **357,** 187-191 (2017).
19  Dean, C. R. *et al.* Boron nitride substrates for high-quality graphene electronics. *Nat. Nanotechnol.* **5,** 722-726 (2010).
20  Woessner, A. *et al.* Highly confined low-loss plasmons in graphene–boron nitride heterostructures. *Nat. Mater.* **14,** 421-425 (2015).
21  Chen, N. *et al*. Boundary-induced excitation of higher-order hyperbolic phonon polaritons. *Nat. Photon.* **19,** 1225-1232 (2025).
22  Giles, A. J. *et al.* Ultralow-loss polaritons in isotopically pure boron nitride. *Nat. Mater.* **17,** 134-139 (2018).
23  Galiffi, E. *et al*. Extreme light confinement and control in low-symmetry phonon-polaritonic crystals. *Nat. Rev. Mater.* **9,** 9-28 (2024).
24  Zheng, C. *et al*. Hyperbolic-to-hyperbolic transition at exceptional Reststrahlen point in rare-earth oxyorthosilicates. *Nat. Commun.* **15,** 7047 (2024).
25  Yves, S. *et al*. Symmetry-driven artificial phononic media. *Nat. Rev. Mater.* **11,** 156-180 (2026).
26  Tan, P. H. *et al*. The shear mode of multilayer graphene. *Nat. Mater.* **11,** 294-300 (2012).
27  Li, P. et al. Infrared hyperbolic metasurface based on nanostructured van der Waals materials. *Science* **359,** 892-896 (2018).
28  Liu, Q. *et al*. Anisotropic Crystallographic Engineering of α-MoO$_3$. *ACS Nano* **19,** 21179-21188 (2025).
29  Hillenbrand, R., Abate, Y., Liu, M., Chen, X. & Basov, D. N. Visible-to-THz near-field nanoscopy. *Nat. Rev. Mater.* **10,** 285-310 (2025).
30  Chen, N. *et al*. Flatland wakes based on leaky hyperbolic polaritons. *Nat. Mater.* **24,** 1569-1575 (2025).
31  Tresguerres-Mata, A. I. F. et al. Directional strong coupling at the nanoscale between hyperbolic polaritons and organic molecules. *Nat. Photon.* **19,** 1196-1202 (2025).
32  Sun, T. *et al*. Van der Waals quaternary oxides for tunable low-loss anisotropic polaritonics. *Nat. Nanotechnol.* **19,** 758-765 (2024).
33  Liu, L. *et al*. Long-range hyperbolic polaritons on a non-hyperbolic crystal surface. *Nature* **644,** 76-82 (2025).

34 Ma, W. *et al*. In-plane anisotropic and ultra-low-loss polaritons in a natural van der Waals crystal. *Nature* **562,** 557-562 (2018).

35 Guan, F. *et al*. Overcoming losses in superlenses with synthetic waves of complex frequency. *Science* **381,** 766-771 (2023).

36 Hu, G. et al. Topological polaritons and photonic magic angles in twisted α-MoO$_3$ bilayers. *Nature* **582,** 209-213 (2020).

37 Chen, M. *et al*. Configurable phonon polaritons in twisted α-MoO$_3$. *Nat. Mater.* **19,** 1307-1311 (2020).

38 Duan, J. *et al*. Multiple and spectrally robust photonic magic angles in reconfigurable α-MoO$_3$ trilayers. *Nat. Mater.* **22,** 867-872 (2023).

39 Ma, X. *et al*. Capillary-Force-Assisted Clean-Stamp Transfer of Two-Dimensional Materials. *Nano Lett.* **17,** 6961-6967 (2017).

40 Huang, G. *et al*. Transfer and beyond: emerging strategies and trends in two-dimensional material device fabrication. *Chem. Soc. Rev.* **55**, 2574-2634 (2026).

41 Chen, Z. *et al*. Broadband measurement of Feibelman's quantum surface response functions. *Proc. Natl Acad. Sci. USA* **122,** e2501121122 (2025).

42 Feibelman, P. J. Surface electromagnetic fields. *Prog. Surf. Sci.* **12,** 287-407 (1982).

43 Yang, Y. *et al*. A general theoretical and experimental framework for nanoscale electromagnetism. *Nature* **576,** 248-252 (2019).

44 Lu, L.-S. *et al*. Layer-Dependent and In-Plane Anisotropic Properties of Low-Temperature Synthesized Few-Layer PdSe$_2$ Single Crystals. *ACS Nano* **14,** 4963-4972 (2020).

45 Oyedele, A. D. *et al*. PdSe$_2$: Pentagonal Two-Dimensional Layers with High Air Stability for Electronics. *J. Am. Chem. Soc.* **139,** 14090-14097 (2017).

46 Anderson, R. L. Germanium-Gallium Arsenide Heterojunctions [Letter to the Editor]. *IBM J. Res. Dev.* **4,** 283-287 (1960).

47 Anderson, R. L. Experiments on Ge-GaAs heterojunctions. *Solid State Electron.* **5,** 341-351 (1962).

48 Ma, X. *et al*. Coherent momentum control of forbidden excitons. *Nat. Commun.* **13,** 6916 (2022).

49 Michaelson, H. B. The work function of the elements and its periodicity. *J. Appl. Phys.* **48,** 4729-4733 (1977).

50 Rut'kov, E. V., Afanas'eva, E. Y. & Gall, N. R. Graphene and graphite work function depending on layer number on Re. *Diamond Relat. Mater.* **101,** 107576 (2020).

51 Álvarez-Pérez, G. et al. Infrared Permittivity of the Biaxial van der Waals Semiconductor α-MoO$_3$ from Near- and Far-Field Correlative Studies. *Adv. Mater.* **32,** 1908176 (2020).

52 Henkelman, G., Arnaldsson, A. & Jónsson, H. A fast and robust algorithm for Bader decomposition of charge density. *Comput. Mater. Sci.* **36,** 354-360 (2006).

53 Kresse, G. & Furthmüller, J. Efficient iterative schemes for ab initio total-energy calculations using a plane-wave basis set. *Phys. Rev. B* **54,** 11169-11186 (1996).

54 Dudarev, S. L., Botton, G. A., Savrasov, S. Y., Humphreys, C. J. & Sutton, A. P. Electron-energy-loss spectra and the structural stability of nickel oxide: An LSDA+U study. *Phys. Rev. B* **57,** 1505-1509 (1998).

55 Kresse, G. & Joubert, D. From ultrasoft pseudopotentials to the projector augmented-wave method. *Phys. Rev. B* **59,** 1758-1775 (1999).

56 Perdew, J. P., Burke, K. & Ernzerhof, M. Generalized Gradient Approximation Made Simple. *Phys. Rev. Lett.* **77,** 3865-3868 (1996).

57 Grimme, S., Antony, J., Ehrlich, S. & Krieg, H. A consistent and accurate ab initio parametrization of density functional dispersion correction (DFT-D) for the 94 elements H-Pu. *J. Chem. Phys.* **132,** 154104 (2010).